\documentclass{article}
\usepackage[utf8]{inputenc}
\usepackage{times}
\usepackage{amsfonts}
\usepackage{amsmath,amssymb}
\usepackage{indentfirst}
\usepackage{wrapfig}
\usepackage{graphicx}
\usepackage{graphics}
\usepackage{caption}
\usepackage{subcaption}
\usepackage[toc,page]{appendix}
\usepackage{color}
\usepackage{cite}
\usepackage[inactive]{srcltx}
\usepackage[T1]{fontenc}
\usepackage{float}
\usepackage{hyperref}
\newcommand{\sech}{\textrm{sech}}
\begin{document}
	\date{}
\begin{center}
	{\Large\bf Quantum state truncation using an optical parametric amplifier and a beamsplitter}
\end{center}
\begin{center}
	{\normalsize E.P. Mattos and A. Vidiella-Barranco \footnote{vidiella@ifi.unicamp.br}}
\end{center}
\begin{center}
	{\normalsize{Gleb Wataghin Institute of Physics - University of Campinas}}\\
	{\normalsize{ 13083-859   Campinas,  SP,  Brazil}}\\
\end{center}
\begin{abstract}
We present a scheme of quantum state truncation in the Fock basis (quantum scissors), 
based on the combined action of a nondegenerate optical parametric amplifier and a 
beamsplitter. Differently from previously proposed linear-optics-based quantum scissors 
devices, which depend on reliable Fock states sources, our scheme requires 
only readily available Gaussian states, such as coherent states inputs (vacuum
state included). A truncated state is generated after performing photodetections in 
the global state. We find that, depending on which output ports each of the two 
photodetectors is positioned, different types of truncated states may be 
produced: i) states having a maximum Fock number of $N$, or ii) states having 
a minimum Fock number $N$. In order to illustrate our method, we discuss 
an example having as input states a coherent state in the beamsplitter and 
vacuum states in the amplifier, and show that the resulting truncated states 
display nonclassical properties, such as sub-Poissonian statistics and 
squeezing. We quantify the nonclassicality degree of the generated states using
the Wigner-Yanase skew information measure. For complementarity, we discuss the efficiency 
of the protocol, e.g., generation probability as well as the effects of imperfections such 
as the detector's quantum efficiency and dark counts rate.
\end{abstract}
%
\section{Introduction}
The engineering of quantum states of light has experienced extraordinary progress in recent years \cite{dodonov02}.
Despite the fact that the photon concept emerged in the early days of quantum theory, the generation of 
pure photon number states, or Fock states $|n\rangle$, has been particularly challenging. Early attempts to generate,
for instance, single photon states to some degree of control occurred only in the 1970s, using nonlinear media, 
\cite{weinberg70} or in atomic systems \cite{clauser74}. This was well after the generation of coherent states 
(laser light) in the 1960s \cite{maiman60}, and before the successful production of squeezed states of light 
\cite{slusher85}. Since those pioneeeing experiments, there have been considerable efforts to generate states of 
light having diverse nonclassical properties \cite{dodonov02,pathak18}, also because they are essential resources 
for the development of quantum technologies \cite{pathak18,browne17,barnett17}. Amongst the proposed methods, we may cite 
quantum state engineering schemes using arrays of beamsplitters with injection of suitable states (usually Fock states) 
followed by photodetections \cite{welsch99}. There are also proposals based on the use of specifically engineered 
nonlinear media \cite{kilin95,avb98}, as well as in cavity QED systems \cite{schleich93,dantsker94,eberly96}. 
Needless to say it is worth looking for alternative generation schemes, since quantum states engineering protocols 
are in general not easy to implement.

An appealing approach is to try to modify an already existing state of light by applying some kind of
operation on it. As examples of such operations we could cite: photon addition \cite{agarwal91}, 
photon subtraction \cite{agarwal92}, and the removal of specific components in the Fock basis (``hole burning") 
\cite{baseia04}. We remark that the removal of the vacuum state is enough to transform an arbitrary state into a 
nonclassical one, as discussed in \cite{lee95}. Another interesting procedure is the so-called quantum state 
truncation, also know as ``quantum scissors" after reference \cite{barnett98}.
A quantum scissors device transforms a quantum state of light, say $|\varphi\rangle$, into a state having a 
finite number of Fock components, that can be a superposition of the vacuum state $|0\rangle$ and the one
photon state $|1\rangle$: $\hat{T}|\varphi\rangle = c_0|0\rangle + c_1|1\rangle$. We note that
the truncation of quantum states in the Fock basis can also be performed in vibrational states 
of a trapped ion system \cite{tombesi00}.
A typical quantum scissors device for light \cite{barnett98} consists of two beamsplitters placed side by side,
having the vacuum state $|0\rangle$ and a single photon Fock state $|1\rangle$ as input states of the first 
beamsplitter, and an arbitrary state $|\varphi\rangle$ as input state of the second beamsplitter. The other
input of the second beamsplitter is precisely the transmitted output of the first beamsplitter. 
Two photodetectors are placed in the output ports of the second beamsplitter, and the detection of 
one photon in one and no photons in the other projects the reflected output of the first beamsplitter into 
a truncated state \cite{barnett98}. Such a process is allowed because the quantum state
$|\varphi\rangle$ is mixed with an entangled state (involving the vacuum state $|0\rangle$ and a 
single photon state $|1\rangle$) in the second beamsplitter. We stress that in general, quantum scissors 
schemes require the injection of Fock states \cite{barnett98,imoto01,ralph20}, i.e., the prior generation of a highly 
nonclassical state. In recent years, there has been renewed interest in the study of quantum state truncation. 
In particular, one can find in the literature a number of works about possible applications, such as: 
entanglement improvement \cite{liao18,zubairy19}, continuous variable quantum key distribution \cite{razavi20}, 
quantum repeaters, \cite{guha20}, and noiseless amplification \cite{ralph20,malaney21}.

Here we propose a hybrid quantum scissors scheme employing linear (beamsplitter) and nonlinear (nondegenerate 
optical parametric amplifier) devices, as displayed in Fig. \ref{parampbs}. We show that, in our method,
truncated states may be generated in a straightforward way without having to resort to Fock states as inputs. 
Rather, it is sufficient to have vacuum states entering the amplifier input ports and a coherent state as input 
to the beamsplitter. We also show that it is possible to generate two distinct classes of truncated states simply 
by placing the photodetectors in different exit ports. Furthermore, our alternative scheme offers additional 
possibilities for the output states, depending on the strength and phase of the parametric amplifier.

This paper is organized as follows: In Section \ref{model} we present our nonlinear-linear quantum scissors. In
Section \ref{example} we study a specific example of truncated state generation using Gaussian states inputs.
We also discuss some nonclassical properties as well as the degree of nonclassicality of the generated states 
using the Wigner-Yanase skew information \cite{zhang19,wigner63}. The efficiency of the protocol taking into 
account imperfections in the photodetections is analysed in Section \ref{eff}, and in Section \ref{conclusion} 
we conclude our work.

\section{A scheme for generalized quantum state truncation}\label{model}

Our proposal is based in sequential interactions using the setup shown in Fig.\ref{parampbs}. It employs 
a nonlinear device, namely a nondegenerate optical parametric amplifier placed besides a linear device, a 
beamsplitter, in such a way that one of the amplifier's output ports (along mode $\hat{b}$, see Fig. \ref{parampbs}) 
becomes one of the beamsplitter's input modes. The other input port of the beamsplitter (mode $\hat{c}$) is fed by 
an arbitrary quantum state of light. Photodetectors may be placed in two output ports, out of the three 
existing ones. We consider two configurations: i) both photodetectors in the two output ports of the beamsplitter 
($\hat{b}_{out}$ and $\hat{c}_{out}$ modes), or ii) one photodetector in an output port of the beamsplitter 
($\hat{c}_{out}$ mode) and the other in the remaining output port of the amplifier ($\hat{a}_{out}$ mode). 
Naturally, the generated quantum state of light, conditioned to the corresponding photodetections, 
will come out through the port which has been left open. As we are going to see below, different families of 
nonclassical states can be generated, depending on the positions in which the detectors are placed. 

We consider a simple case, in which the initial state entering the amplifier/beamsplitter device is such 
that both input modes of the amplifier are in the vacuum state, and the input mode $\hat{c}$ 
(beamsplitter) is in a generic pure state $|\psi\rangle = \sum \psi_i |i\rangle$, or 
\begin{equation}
|\Psi_{in}\rangle=\sum_{i=0}^\infty\psi_i|0, 0, i\rangle,
\end{equation}
where $|0, 0, i\rangle \equiv |0\rangle_a \otimes |0\rangle_b \otimes |i\rangle_c$.
The combined action of the amplifier and beamsplitter 
on the initial state $|\Psi_{in}\rangle$ may be represented as
\begin{equation}
	|\Phi_{out}\rangle=\hat{R}(\theta)\hat{S}(\xi)|\Psi_{in}\rangle,
\end{equation}
being $\hat{S} = \exp\left[\xi^* \hat{a}\hat{b} - \xi \hat{a}^\dagger\hat{b}^\dagger \right]$ the two-mode 
squeezing operator and $\hat{R} = \exp\left[i\theta(\hat{a}^\dagger\hat{b} + \hat{a}\hat{b}^\dagger)\right]$ 
the operator associated to the beamsplitter action. The relevant parameters here are $\xi = s e^{i\phi}$, where 
$s$ is basically the strength of the amplifier ($s \geq 0$), and $\phi$ is the phase of the pump field (treated as
classical here). The parameter $\theta$ is related to the (complex) transmittance $\mathtt{T}$ and reflectance 
$\mathtt{R}$ of the beamsplitter $(|\mathtt{T}|^2 + |\mathtt{R}|^2 = 1)$.
After some algebra (see details in the Appendix), the output state prior to the photodetections will read
\begin{eqnarray}
	\label{generalstate}
	|\Phi_{out}\rangle&=&\sum_{i=0}^\infty\sum_{n=0}^\infty\sum_{j=0}^i\sum_{m=0}^n\frac{\sqrt{i!}}{j!(i-j)!}
	\frac{\sqrt{n!}}{m!(n-m)!}\sqrt{(i-j+n-m)!} \sqrt{(j+m)!}\,\psi_i\, A_n(s,\phi) \nonumber \\
	&\times& \mathtt{T}^j \mathtt{T}^{*n-m} 
	\mathtt{R}^m(-\mathtt{R}^*)^{i-j}|n, i-j+n-m, j+m\rangle, 
\end{eqnarray}
with $A_n(s,\phi)=\sech\,s(-e^{i\phi}\tanh{s})^n$.

\subsection{Photon detectors placed in both the beamsplitter's output ports}\label{case-a}

If the photodetectors are placed in such a way that one is at $\hat{b}_{out}$ and the other at 
$\hat{c}_{out}$, having one photon detected in $\hat{b}_{out}$ and no photons detected in $\hat{c}_{out}$, 
the following conditional truncated state will be generated in mode $\hat{a}_{out}$

\begin{equation}
|\Phi_{a}^{(1,0)}\rangle	= -\frac{1}{\sqrt{p_{a}^{(1,0)}}}\sech\,s
\left(\psi_1 \mathtt{R}^*|0\rangle + \psi_0e^{i\phi}\tanh{s}\ \mathtt{T}^*|1\rangle\right),\label{truncated1}
\end{equation}
with
\begin{equation}
	p_{a}^{(1,0)} = \sech^2s\left(|\mathtt{R}|^2 |\psi_1|^2 + |\mathtt{T}|^2 |\psi_0|^2 \tanh^2s \right).
	\label{eq:probgena}
\end{equation}
The state in Eq. (\ref{truncated1}), a quantum superposition of the vacuum state and the one-photon state, has 
the form of a typical truncated state obtained via a conventional quantum scissors device \cite{barnett98}. 
We remind the reader that in the quantum scissors schemes previously discussed in the literature, 
states which are difficult to generate in a controlled way (Fock states) are required as input states. 
In our method there is no need of previous generation of Fock states; the input states in the amplifier 
are simply vacuum states (apart from the classical pump), with an arbitrary state $|\psi\rangle$ entering 
the beamsplitter's port $\hat{c}$.

It is possible to generalize the result above for $N$ photons being detected in $\hat{b}_{out}$ and no photons in
$\hat{c}_{out}$. In this case the generated state will be

\begin{equation}
\label{truncatedn}
|\Phi_{a}^{(N,0)}\rangle = \frac{1}{\sqrt{p_{a}^{(N,0)}}}\sech\,s\sum_{n=0}^N\sqrt{\frac{N!}{n!(N-n)!}}
\psi_{N-n}(-e^{i\phi}\tanh\,s)^n \mathtt{T}^{*n}(-\mathtt{R}^*)^{N-n}|n\rangle,
\end{equation}
with
\begin{equation}
	p_{a}^{(N,0)} = \sech^2s\sum_{n=0}^N\frac{N!}{n!(N-n)!}|\psi_{N-n}|^2
	|\mathtt{T}|^{2n}|\mathtt{R}|^{2(N-n)} \tanh^{2n}s.\label{eq:probgenan}
\end{equation}
In other words, our scheme allows, in principle, the generation of a truncated state up to Fock number $N$. 

We note that Fock states as well as the vacuum state can also be output states of the 
quantum scissors device for particular values of the transmittance. From Eq. (\ref{truncated1}) it follows
that if $|\mathtt{T}| = 1$ a one photon state $|1\rangle$ is generated, while if $|\mathtt{T}| = 0$, the
resulting state in $\hat{a}_{out}$ will be simply the vacuum state.

\subsection{Photon detectors placed in the beamsplitter's and amplifier's output ports}\label{case-b}

In this case the photodetectors will be placed in the ports corresponding to the $\hat{a}_{out}$ (amplifier) and
$\hat{c}_{out}$ (beamsplitter) modes. For instance, if one photon is recorded at the amplifier output and no
photon is detected at the beamsplitter output, the generated state at port $\hat{b}_{out}$ will be

\begin{equation}
|\Phi_{b}^{(1,0)}\rangle =\frac{1}{\sqrt{p_{b}^{(1,0)}}}\sech\,s(-e^{i\phi}\tanh\,s)\mathtt{T}^* 
\sum_{i=0}^\infty\sqrt{i+1}\psi_i(-\mathtt{R}^*)^i|i+1\rangle, \label{vacuumremoved}
\end{equation}

with
\begin{equation}
	p_{b}^{(1,0)} =\sech^2s \tanh^2s\ |\mathtt{T}|^2\sum_{i=0}^\infty(i+1)|\psi_i|^2|\mathtt{R}|^{2i}.
\end{equation}

Note that the vacuum component $|0\rangle$ has been removed from the state in Eq. (\ref{vacuumremoved}), i.e.,
the states of the type $|\Phi_{b}^{(1,0)}\rangle$ are nonclassical \cite{lee95}.

Again, we may generalize the above result if $N$ photons are recorded at the amplifier's output port 
(mode $\hat{a}_{out}$) and no photon is recorded at the beamsplitter port. In this case, the generated 
state will read
\begin{equation}
|\Phi_{b}^{(N,0)}\rangle = \frac{1}{\sqrt{p_{b}^{(N,0)}}}\sech\, s(-e^{i\phi}
\tanh\, s)^N \mathtt{T}^{*N}
\sum_{i=0}^\infty\sqrt{\frac{(i+N)!}{i!N!}}\psi_i(-\mathtt{R}^*)^i|i+N\rangle
\end{equation}
with
\begin{equation}
p_{b}^{(N,0)} = \sech^2s \tanh^{2N}s\ |\mathtt{T}|^{2N}\sum_{i=0}^\infty\frac{(i+N)!}{i!N!}|\psi_i|^2|\mathtt{R}|^{2i}.
\end{equation}
Thus, such scheme makes possible to generate states truncated from Fock number $N$, i.e., all
components having $n < N$ being null.

Interestingly, the two different photodetectors placements discussed lead to the generation of states 
that are somehow ``complementary": in Section \ref{case-a} we showed how states with a maximum Fock number 
$N$ can be generated, and here in Section \ref{case-b}, we saw that it is also possible to generate states having a 
minimum Fock number $N$.

\section{State generation from coherent states: nonclassical properties}\label{example}

We may use the parametrization $\mathtt{T} = \cos\theta$ and $\mathtt{R} = i\sin\theta$, so that we are 
left with three parameters: $(s,\phi)$, which are related to the amplifier/pump, and $\theta$, related to 
the beamsplitter's transmittance. This gives a great flexibility to our generation scheme, since we can 
tune the properties of the generated states by changing experimentally controlled parameters.
Now we would like to illustrate our method by choosing specific input states (mode $\hat{c}$), namely, the 
``quasi-classical" coherent states $|\psi\rangle = |\alpha\rangle$, with $\alpha = |\alpha| e^{i\beta}$. 
In this case the truncated states $|\Phi_{a}^{(N,0)}\rangle$ (and their properties), will depend on
the phase difference $\phi - \beta$, as one can see in Eq. (\ref{truncatedn}).

\subsection{Sub-Poissonian statistics}

\begin{figure}[h]
	\centering
	\includegraphics[scale=0.25]{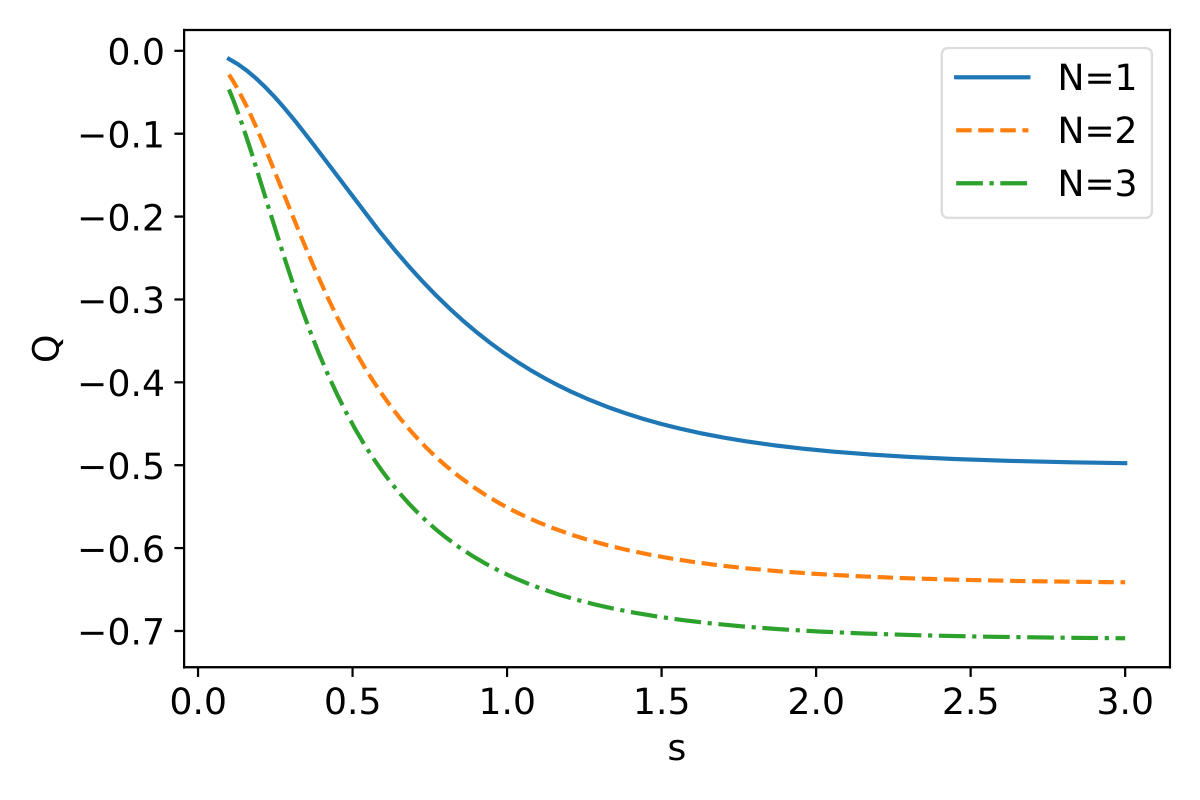}
	\caption{Mandel's $Q$ parameter relative to state $|\Phi_a\rangle$ as a function of $s$ 
		for $\phi - \beta = \pi/2$ rad and $\theta=\pi/4$ rad.}
	\label{asqp}
\end{figure}

\begin{figure}[h]
	\centering
	\includegraphics[scale=0.7]{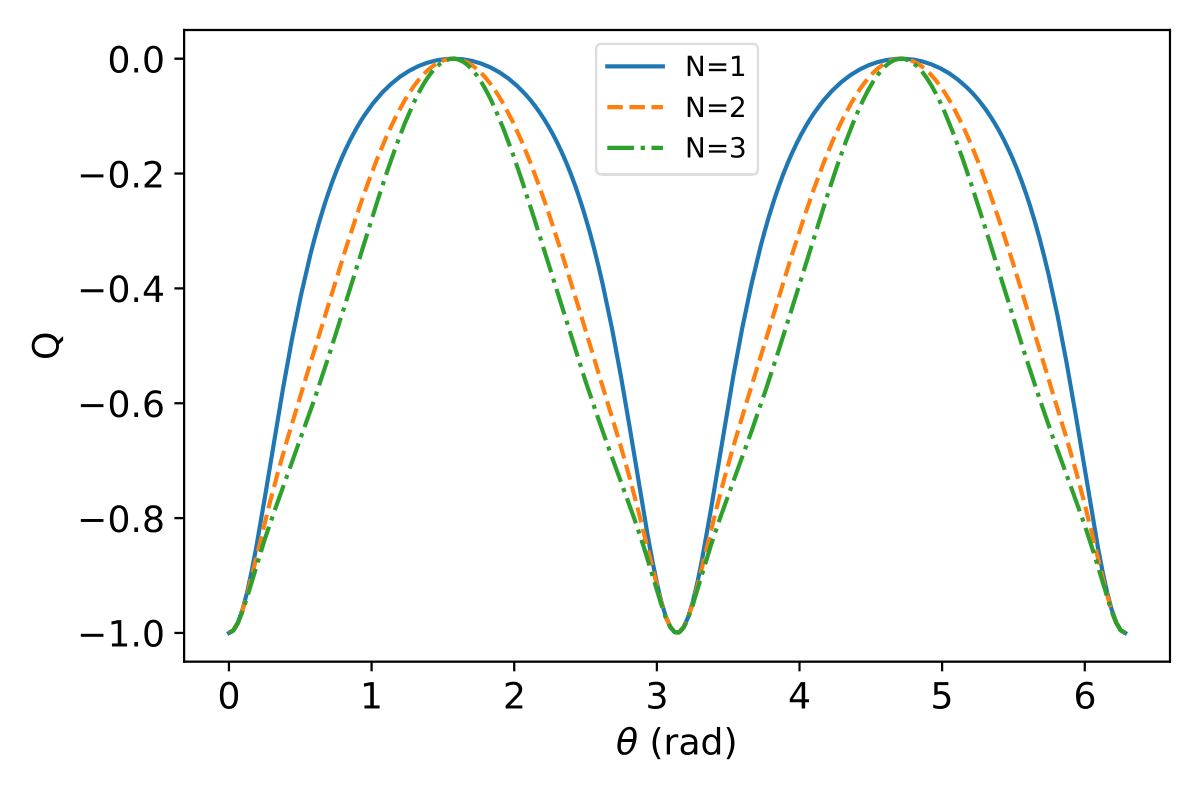}
	\caption{Mandel's $Q$ parameter relative to state $|\Phi_a\rangle$ as a function of $\theta$ 
		for $s = 0.5$ and  $\phi - \beta =\pi/2$ rad.}
	\label{atq}
\end{figure}

A well-known measure of photon number fluctuations is the Mandel $Q$ parameter, 
defined as $Q =  \left\langle (\Delta \hat{n})^2\right\rangle/\left\langle \hat{n} \right\rangle - 1$.
It has a minimum value of $Q = -1$ for Fock states, and is null for coherent states, i.e., it
indicates deviations from the characteristic Poissonian photon statistics of a coherent state.
We firstly analyze the occurrence of sub-Poissonian statistics of the generated truncated state 
$|\Phi_a\rangle$ discussed in Section \ref{case-a}. To begin with, we may set $|\alpha| = 1$, 
$\phi - \beta = \pi/2$ rad, $\theta = \pi/4$ rad (a $50:50$ beamsplitter) and vary the strength $s$. 
The result, in Fig. \ref{asqp} (for different values of $N$), shows that the generated states exhibit 
sub-Poissonian statistics.
If we now set the parameters $s = 0.5$, $\phi - \beta = \pi/2$ rad and vary $\theta$, we obtain the 
results shown in Fig. \ref{atq}, i.e., the generated states are also mostly sub-Poissonian.
The states discussed in Section \ref{case-b}, $|\Phi_b\rangle$,  may also exhibit sub-Poissonian 
statistics, although in a lesser degree than the states $|\Phi_a\rangle$.

\begin{figure}[h]
	\centering
	\includegraphics[scale=0.8]{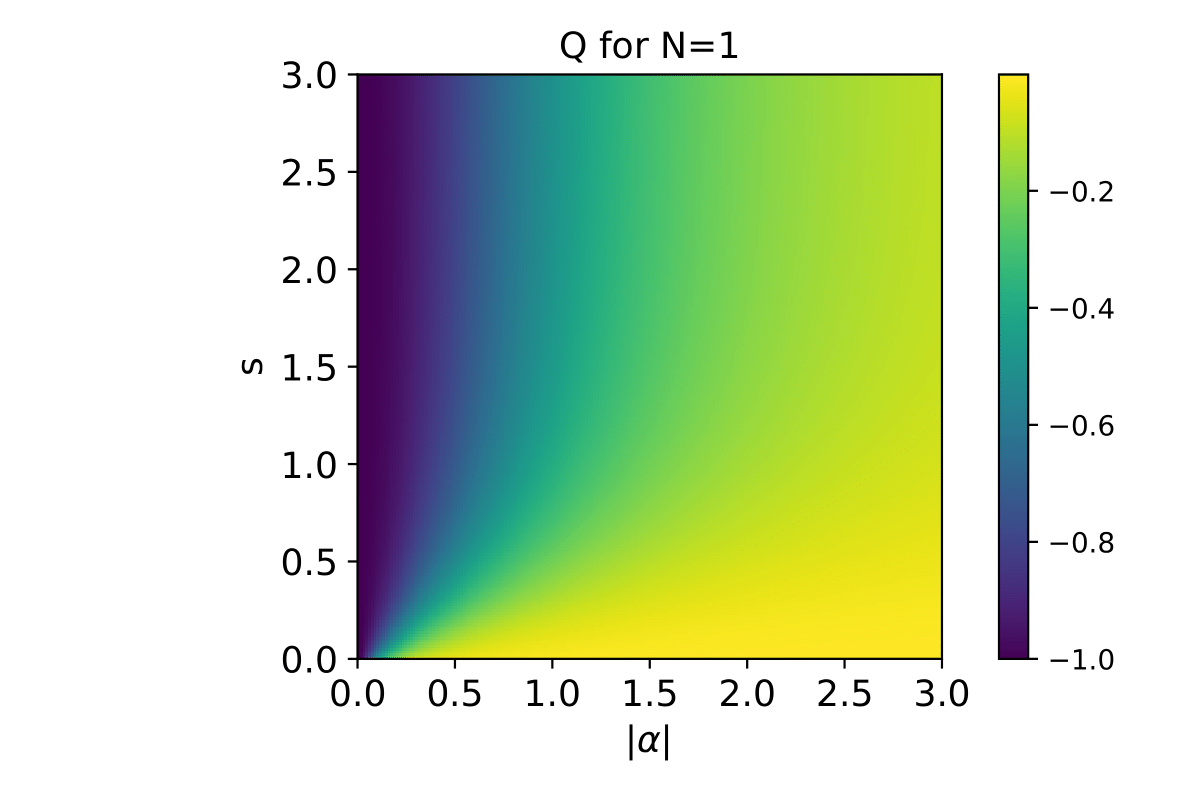}
	\caption{Mandel's $Q$ parameter relative to state $|\Phi_a\rangle$ as a function of $|\alpha|$ 
		and $s$, for  $\theta=\pi/4$ rad, $\phi - \beta=\pi/2$ rad, and $N = 1$.}
	\label{q2d}
\end{figure}

In our scheme, energy is injected into the system via both the classical pump and the input field 
(mode $\hat{c}$). We therefore expect that the nonclassical properties of the output field will depend on $s$, 
as well as on $\alpha$ (in case of a coherent state input). In Fig. \ref{q2d} we have a plot of Mandel's
$Q$ parameter as a function of $s$ and $|\alpha|$, for the state $|\Phi_a\rangle$ with $N = 1$. 
We note that for larger values of $|\alpha|$ the output state is driven onto a Poissonian state. 
Yet, small values of $|\alpha|$ combined with a not too weak pumping favors the generation of 
sub-Poissonian states.

\subsection{Quadrature squeezing}

\begin{figure}[h]
	\centering
	\includegraphics[scale=0.25]{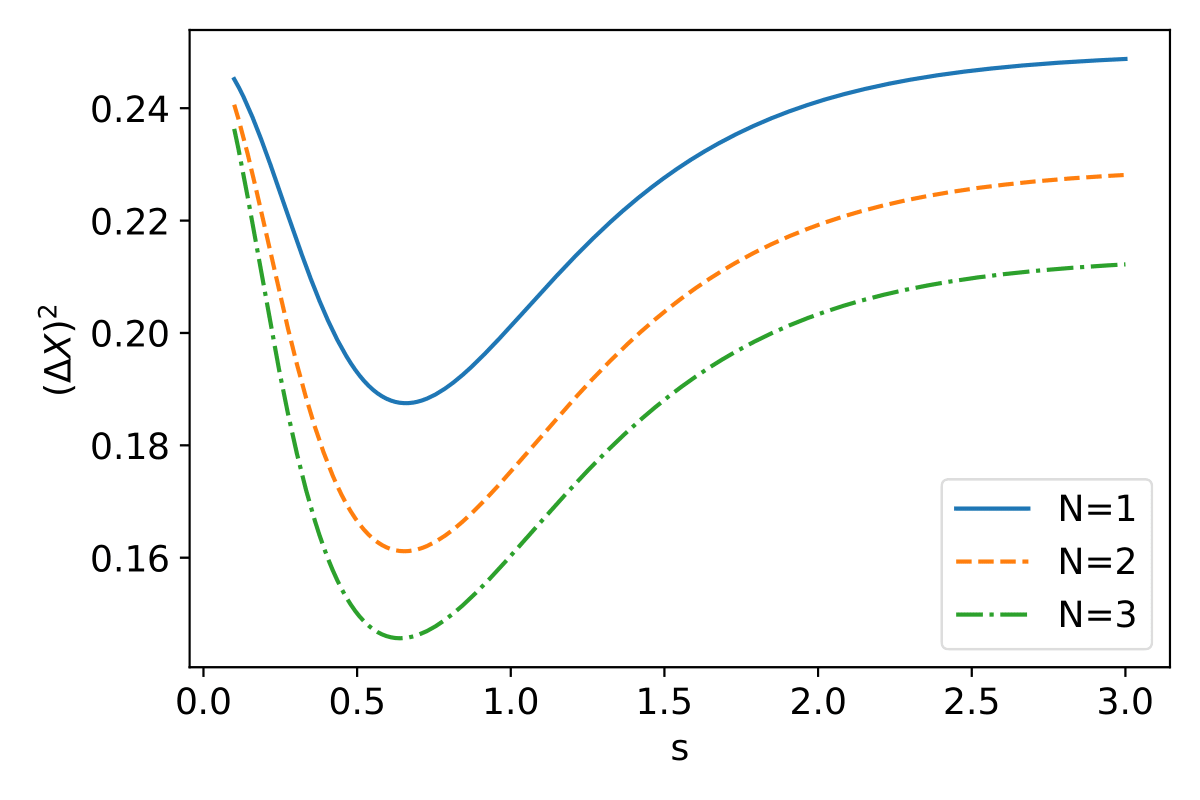}
	\caption{Variance of $\hat{X}$ relative to state $|\Phi_a\rangle$ as a function of $s$ 
		for $\phi - \beta = \pi/2$ rad and $\theta=\pi/4$ rad.}
	\label{asx}
\end{figure}

\begin{figure}[h]
	\centering
	\includegraphics[scale=0.7]{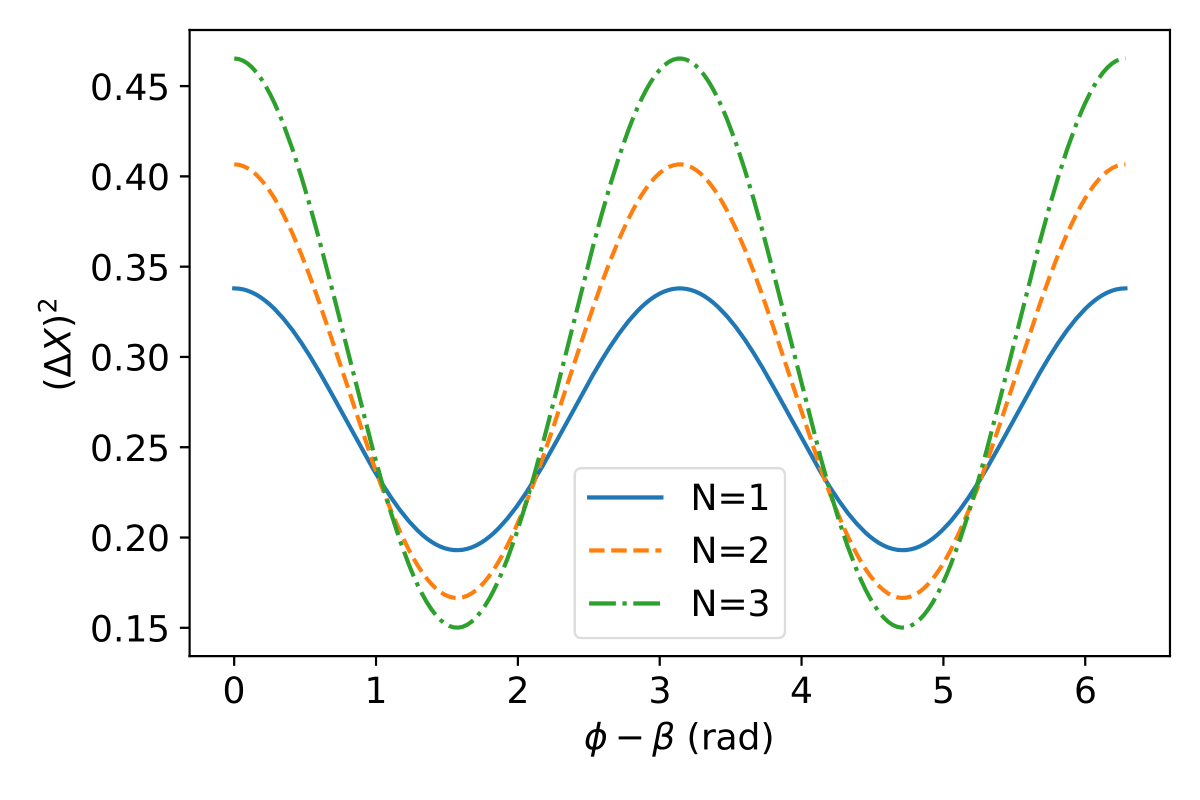}
	\caption{Variance of $\hat{X}$ relative to state $|\Phi_a\rangle$ as a function of $\phi - \beta$
		for $s = 0.5$ and $\theta=\pi/4$ rad.}
	\label{apx}
\end{figure}

\begin{figure}[h]
	\centering
	\includegraphics[scale=0.7]{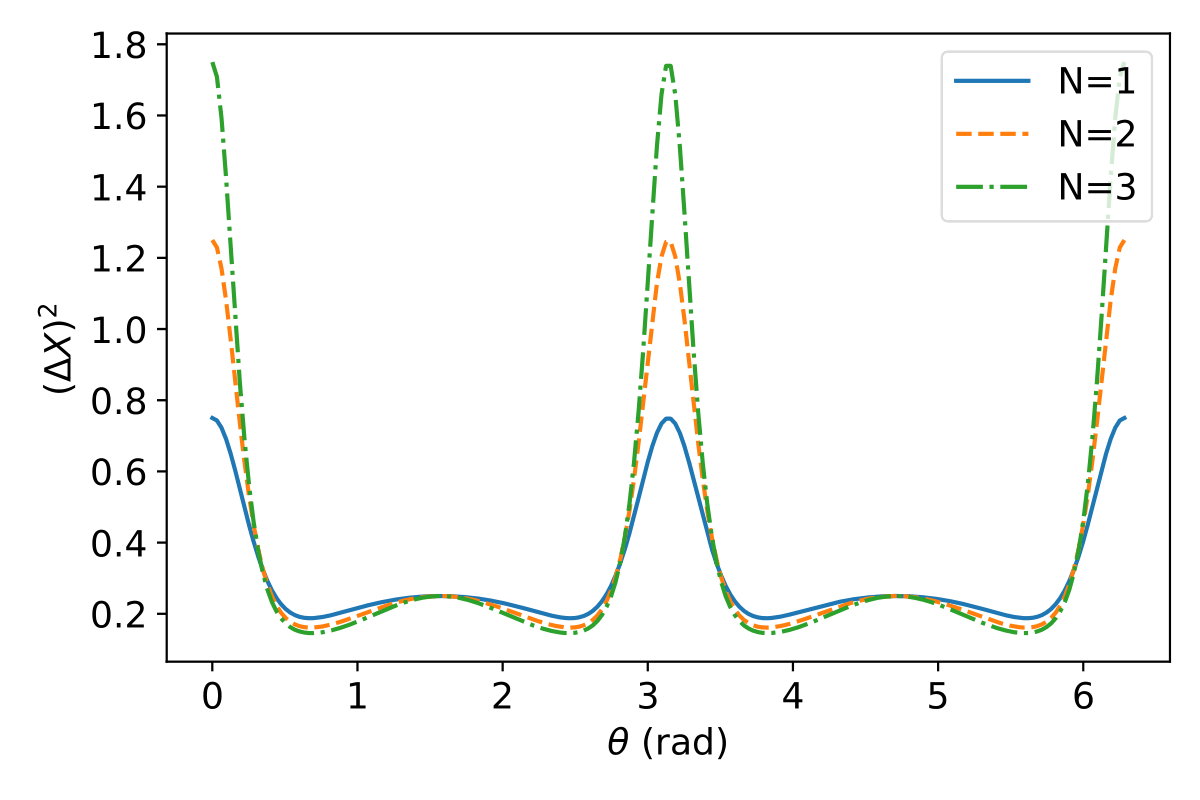}
	\caption{Variance of $\hat{X}$ relative to state $|\Phi_a\rangle$ as a function of $\theta$
		for $s = 0.5$ and $\phi - \beta=\pi/2$ rad.}
	\label{atx}
\end{figure}

Another important nonclassical feature to be discussed is the so-called squeezing; the reduction of 
fluctuations in the quadrature variables below the characteristic value of a coherent state. For instance, 
if $\left\langle (\Delta \hat{X})^2\right\rangle < 1/4$ the quadrature $\hat{X}$, defined as
$\hat{X} = (\hat{a} + \hat{a}^\dagger)/2$, is said to be squeezed. 
The truncated states $|\Phi_a\rangle$ may exhibit squeezing in the $\hat{X}$ quadrature for 
$\phi - \beta = \pi/2$ rad and $\theta = \pi/4$ rad, as shown in Fig. \ref{asx}, where the variance of 
$\hat{X}$ is plotted as a function of $s$. Squeezing may also be present for different combinations of 
the involved parameters, as it is evident from the plots of $\left\langle (\Delta \hat{X})^2\right\rangle$ 
as a function of $\phi - \beta$ (Fig. \ref{apx}), with  $\theta = \pi/4$ rad, as well as
a function of $\theta$ (Fig. \ref{atx}), with $\phi - \beta = \pi/2$ rad. In both cases $s = 0.5$, and 
squeezing occurs for ranges of values of $\phi - \beta$ (or $\theta$). The states $|\Phi_b\rangle$ 
may also exhibit squeezing.

\begin{figure}[h]
	\centering
	\includegraphics[scale=0.8]{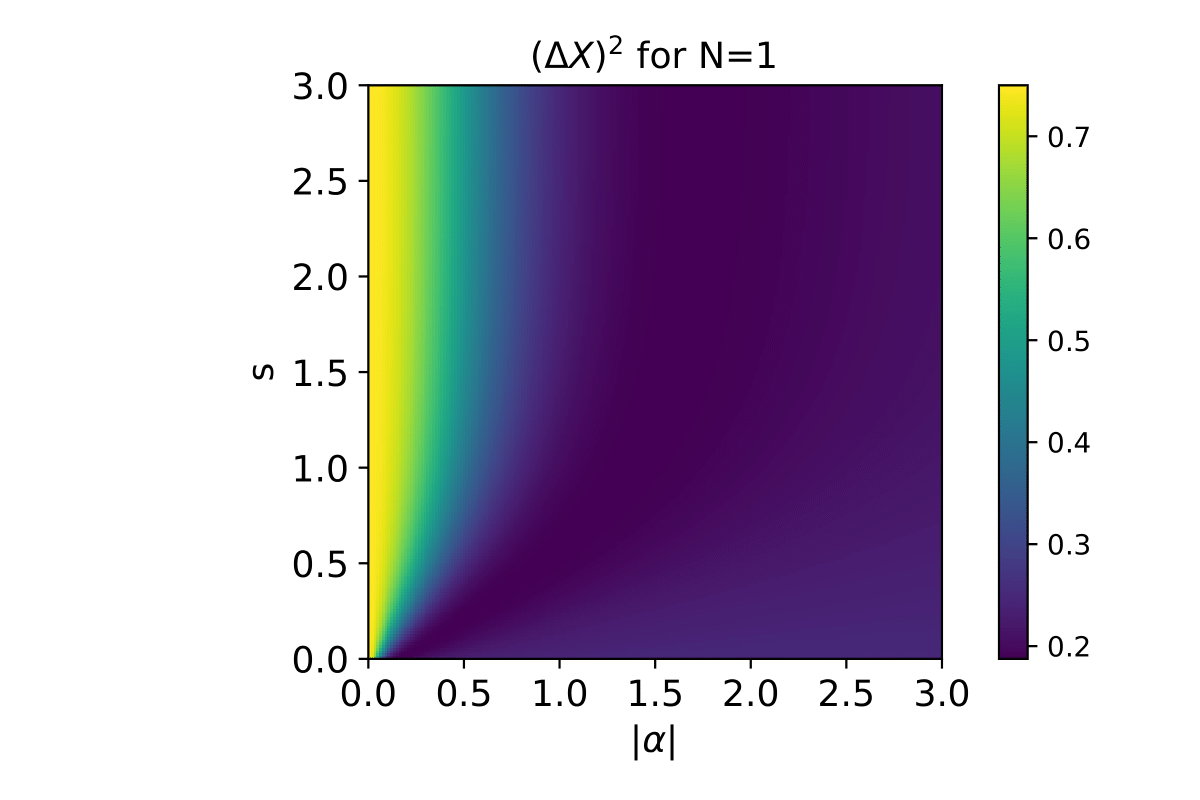}
	\caption{Variance of $\hat{X}$ relative to state $|\Phi_a\rangle$ as a function of $|\alpha|$ 
		and $s$, for  $\theta=\pi/4$ rad, $\phi - \beta=\pi/2$ rad and $N=1$.}
	\label{sq2d}
\end{figure}

Squeezing in the truncated state $|\Phi_a\rangle$ depends not only on the pump strength $s$ 
but also on the input state amplitude, $\alpha$. As clearly shown in Fig. \ref{sq2d}, combinations of
values of $s$ and $|\alpha|$, may yield significant amounts of squeezing to the generated states.

\subsection{Nonclassicality}

Properties such as sub-Poissonian statistics and squeezing capture different nonclassical aspects of
quantum states of light. Nonetheless, due to the multi-sided nature of quantumness (nonclassicality), 
it is not an easy task to find a quantity that would contain as much information as possible about the 
nonclassical character of a quantum state. So far, we have witnessed efforts to quantify nonclassicality from 
different perspectives and, as a consequence, several figures of merit have been introduced for this 
purpose. We may find in the literature works discussing various nonclassicality criteria, e.g., 
distance-based measures \cite{hillery87}, nonclassical depth \cite{lee91}, quadrature-based measures 
\cite{vogel02,vogel05}, negativity of phase space distributions \cite{zyczkowski04} and operator 
ordering sensitivity \cite{horoshko19}. A recently introduced and interesting information-theoretic 
nonclassicality quantifier is the Wigner-Yanase skew information \cite{zhang19,wigner63}. 
For a pure, single mode state of the electromagnetic field 
$|\Psi\rangle$, the skew information is given by \cite{zhang19}
\begin{equation}
W(|\Psi\rangle) = \frac{1}{2} + \langle\Psi| \hat{a}^\dagger\hat{a}|\Psi\rangle - 
	\langle\Psi| \hat{a}^\dagger|\Psi\rangle \langle| \Psi \hat{a}|\Psi\rangle.
\end{equation}

\begin{figure}[h]
	\centering
	\includegraphics[scale=0.25]{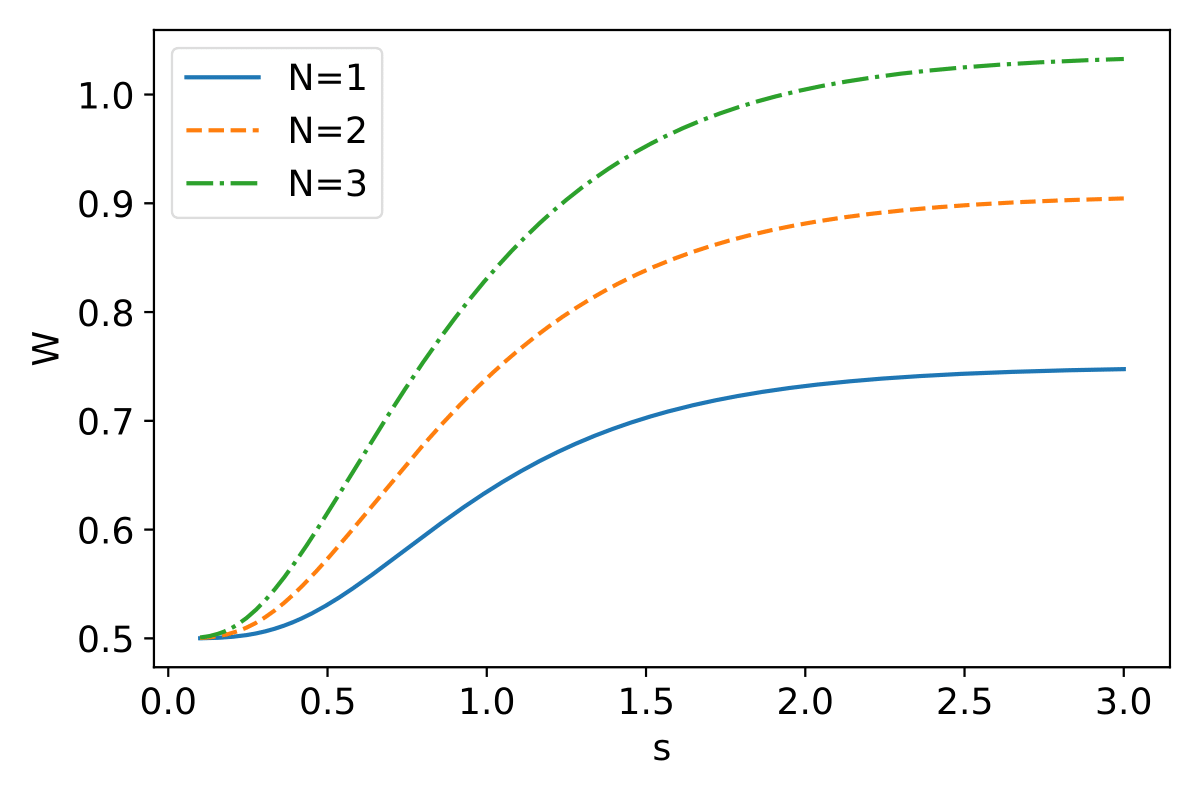}
	\caption{Skew information relative to state $|\Phi_a\rangle$ as a function of $s$ 
		for $\phi - \beta =\pi/2$ rad and $\theta=\pi/4$ rad.}
	\label{asw}
\end{figure}

\begin{figure}[h]
	\centering
	\includegraphics[scale=0.7]{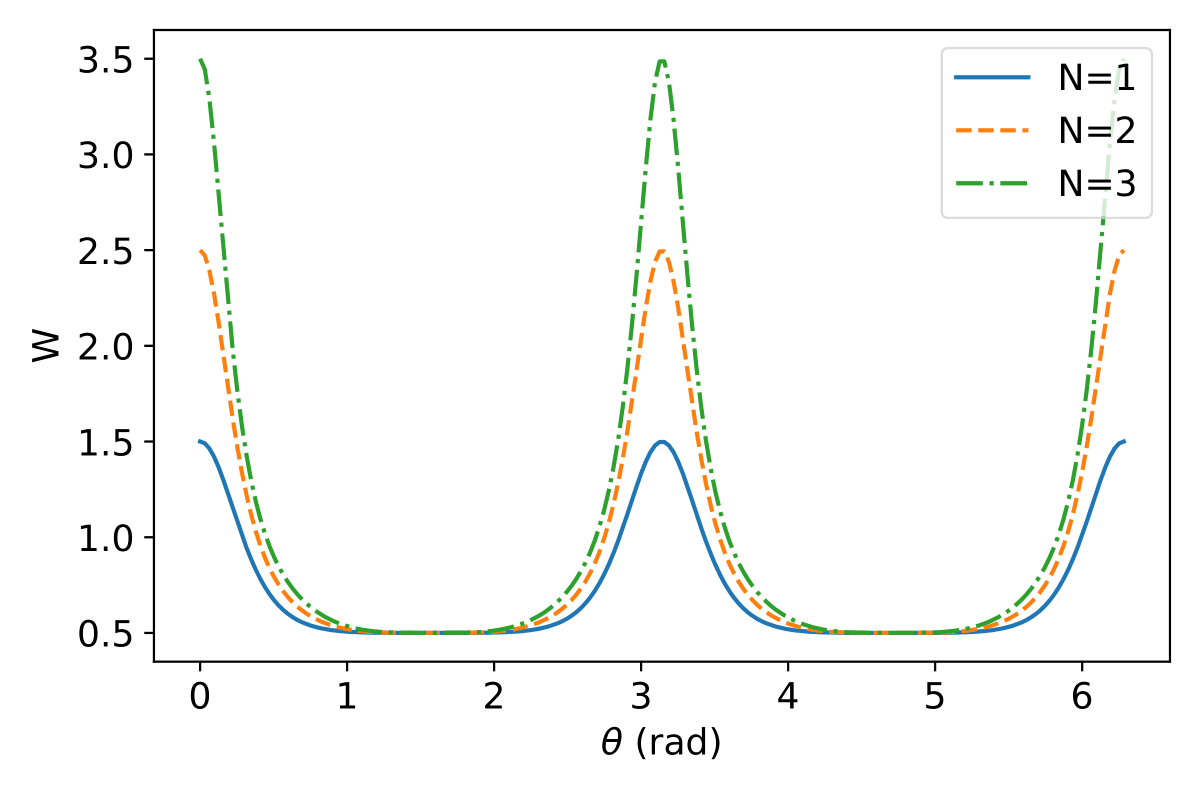}
	\caption{Skew information relative to state $|\Phi_a\rangle$ as a function of $\theta$ 
		for $s = 0.5$ and $\phi - \beta =\pi/2$ rad.}
	\label{atw}
\end{figure}

Among other interesting properties, the skew information is non-negative and for pure
states has a minimum value of $W_{min} = 1/2$ (coherent states).  Also, 
larger values of $W$ indicate a larger nonclassical character of a given state \cite{zhang19}. 
We evaluated the skew information $W$ for the states generated in our scissors device. 
For instance, in Fig. \ref{asw} we have plotted $W$ as a function of $s$ for the states 
$|\Phi_a\rangle$ setting $\phi - \beta = \pi/2$ rad and $\theta = \pi/4$ rad.
We note that the skew information is an increasing function of $s$, which can be 
associated to an increasing sub-Poissonian character (see Fig. \ref{asqp}). Also in this case there 
are variable levels of squeezing, as seen in Fig. \ref{asx}, i.e., $W$ captures an overall nonclassical 
behavior of those states. We could also set $s = 0.5$, $\phi - \beta = \pi/2$ rad and vary $\theta$. 
The resulting plots are shown in Fig. \ref{atw}.
Thus, the Wigner-Yanase skew information captures the nonclassical character of the generated 
states, and it can be associated to sub-Poissonian statistics and/or squeezing. 
Besides, as it is clearly seen in the graphs, the nonclassical character is more pronounced for 
states having a higher maximum Fock number $N$.

\section{Efficiency of the protocol}\label{eff}

Quantum scissors rely on photodetections, and hence perform quantum state truncation 
non-deterministically. Therefore, even under ideal conditions, success probabilities may 
be associated to its realization. Besides, photodetectors are imperfect, which certainly 
has a negative impact on the quality of the generated states. In what follows we are
going to discuss some aspects of the efficiency of our scheme.

\subsection{Generation probabilities}

\begin{figure}[h]
	\centering
	\includegraphics[scale=0.8]{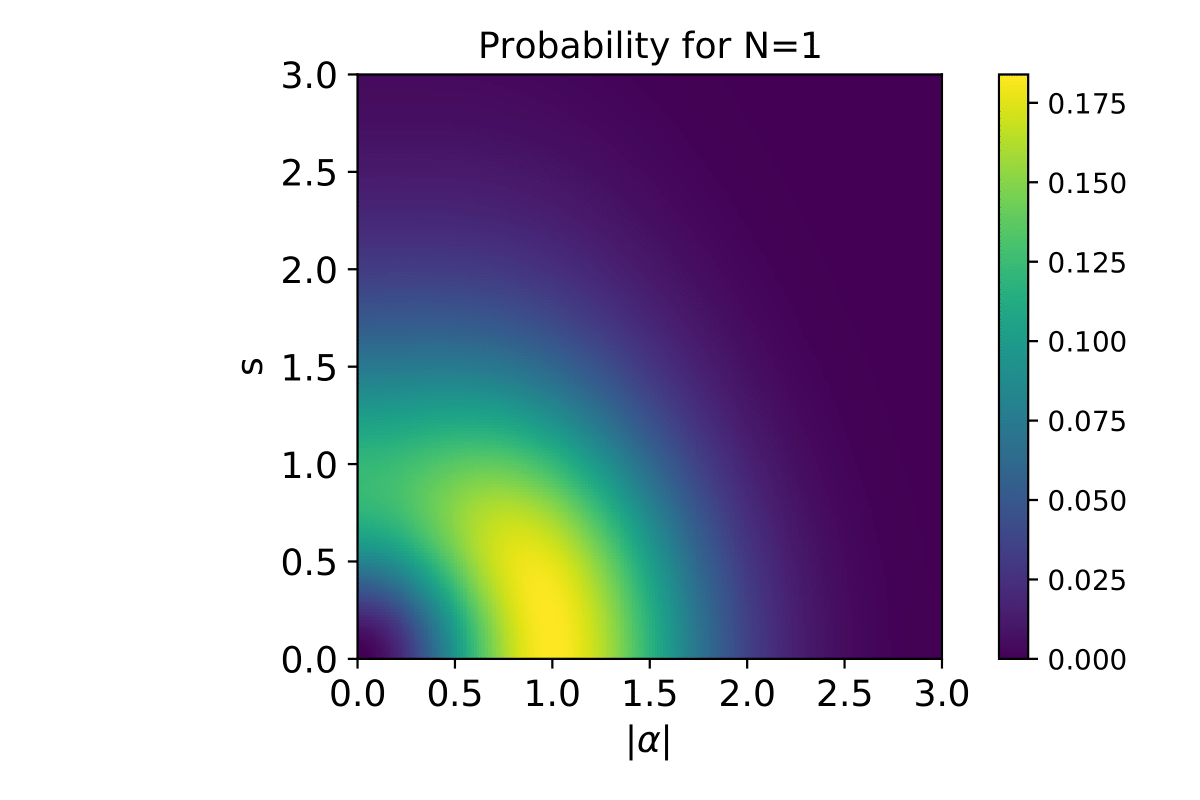}
	\caption{Probability of generation of state $|\Phi_a\rangle$ as a function of $|\alpha|$ 
		and $s$, for $N = 1$, $\theta=\pi/4$ rad and $\phi - \beta=\pi/2$ rad.}
	\label{pa2d1}
\end{figure}

\begin{figure}[h]
	\centering
	\includegraphics[scale=0.8]{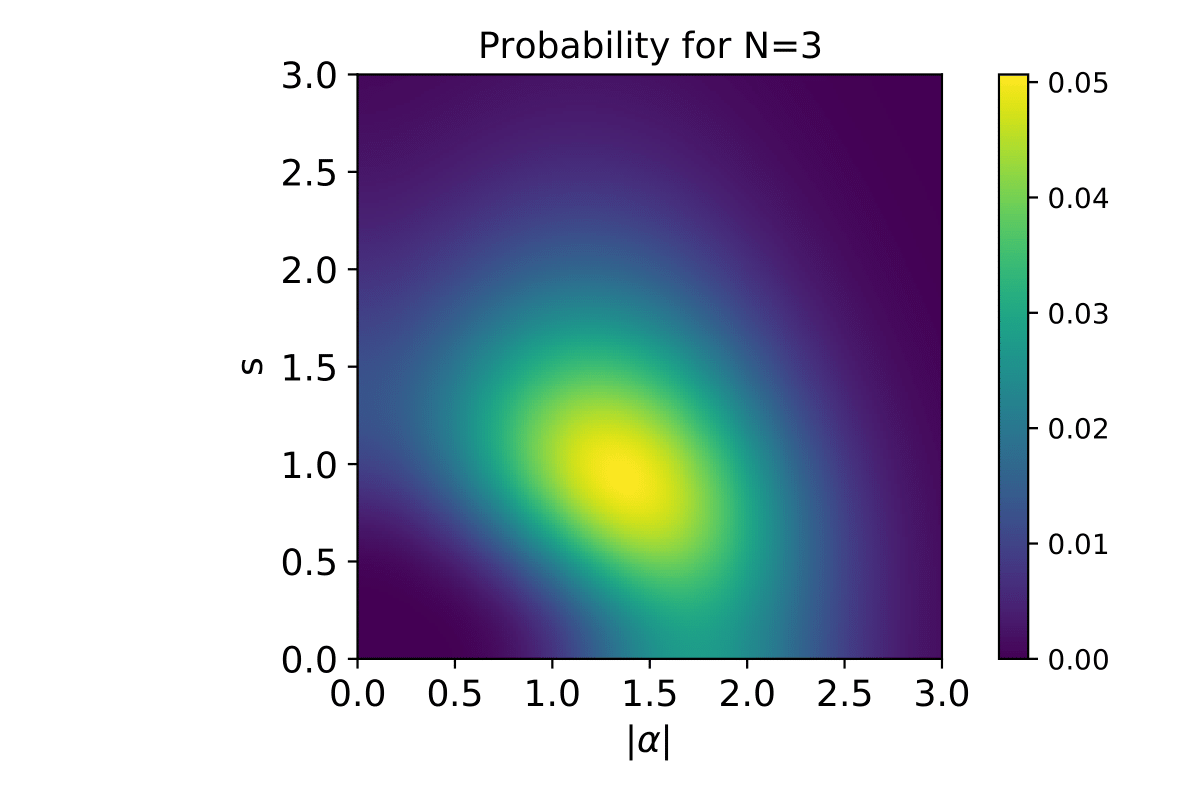}
	\caption{Probability of generation of state $|\Phi_a\rangle$ as a function of $|\alpha|$ 
		and $s$, for $N = 3$, $\theta=\pi/4$ rad and $\phi - \beta=\pi/2$ rad.}
	\label{pa2d3}
\end{figure}

The probability of generation of state $|\Phi_{a}^{(N,0)}\rangle$, that is, $p_{a}^{(N,0)}$, is given by 
Eq.(\ref{eq:probgenan}) in Section \ref{case-a}. It basically depends on $\theta$, $s$ and the 
coefficients $\psi_i$ (actually $|\psi_i|^2$). Again, we consider for simplicity a coherent state 
input $|\alpha\rangle$ and a $50:50$ beamsplitter ($\theta = \pi/4$ rad). We may gauge the dependence of the 
probability of generation on the modulus of the coherent amplitude, $|\alpha|$, as well as the amplifier 
strength $s$ by plotting $p_{a}^{(N,0)}$ as a function of these quantities. This is shown in Fig. \ref{pa2d1} 
for $N = 1$ and in Fig. \ref{pa2d3} for $N = 3$. Clearly there are optimum values of $|\alpha|$ 
and $s$ that maximize $p_{a}^{(N,0)}$ for a given $N$. Naturally, for $N = 3$ there is a 
substantial drop in the probability of generation, compared to $N = 1$, while the maximum value of 
$p_{a}^{(3,0)}$ occurs for slightly larger values of both $|\alpha|$ and $s$, as we see in the figures.

\subsection{Nonideal photodetection}

Despite the advances regarding the quality of photodetectors, those devices are still not perfect. 
Some incoming photons may not be recorded (quantum efficiency is not $100\%$),
and sometimes detectors are spuriously activated (dark counts). We assume that 
photon-number-resolving detectors (PNRD) are employed, and counts up to $N$ photons are feasible \cite{sasaki13}.
The imperfections of a single detector can be suitably modeled via the following POVM, \cite{imoto01,pegg98}
\begin{equation}
\hat{\Pi}_N = \sum_ {n=0}^N \sum_{m=n}^\infty\frac{e^{-\nu}\nu^{N-n}}{(N-n)!}\eta^n(1-\eta)^{m-n}C_n^m|m\rangle\langle m|,
\label{povm}
\end{equation}
where $\eta$ is the detector's quantum efficiency, $\nu$ the dark count probability, and $C_n^m$ are 
binomial coefficients. 
In the setup we are considering here, the action of each one of the detectors will be modeled
by $\hat{\Pi}_0$ and $\hat{\Pi}_N$ for zero and for $N$ photon counts, respectively. We assume the 
same efficiency $\eta$ and dark count rate $\nu$ for both detectors. Due to these imperfections,
the generated field should be represented by a density operator, calculated by tracing over the detected modes.
If the detectors are placed in output ports $\hat{b}_{out}$ and $\hat{c}_{out}$, the resulting 
state will be,
\begin{equation}
	\hat{\rho}^{(\eta,\nu)}_a = {\cal A}\,\mbox{Tr}_{b,c}\left[\hat{\Pi}_0 \hat{\Pi}_N |\Phi_{out}\rangle\langle\Phi_{out}|\right],
\end{equation}
where $|\Phi_{out}\rangle$ is the state in Eq. (\ref{generalstate}) and ${\cal A}$ is a normalizing constant.

\begin{figure}
	\centering
	\includegraphics[scale=0.7]{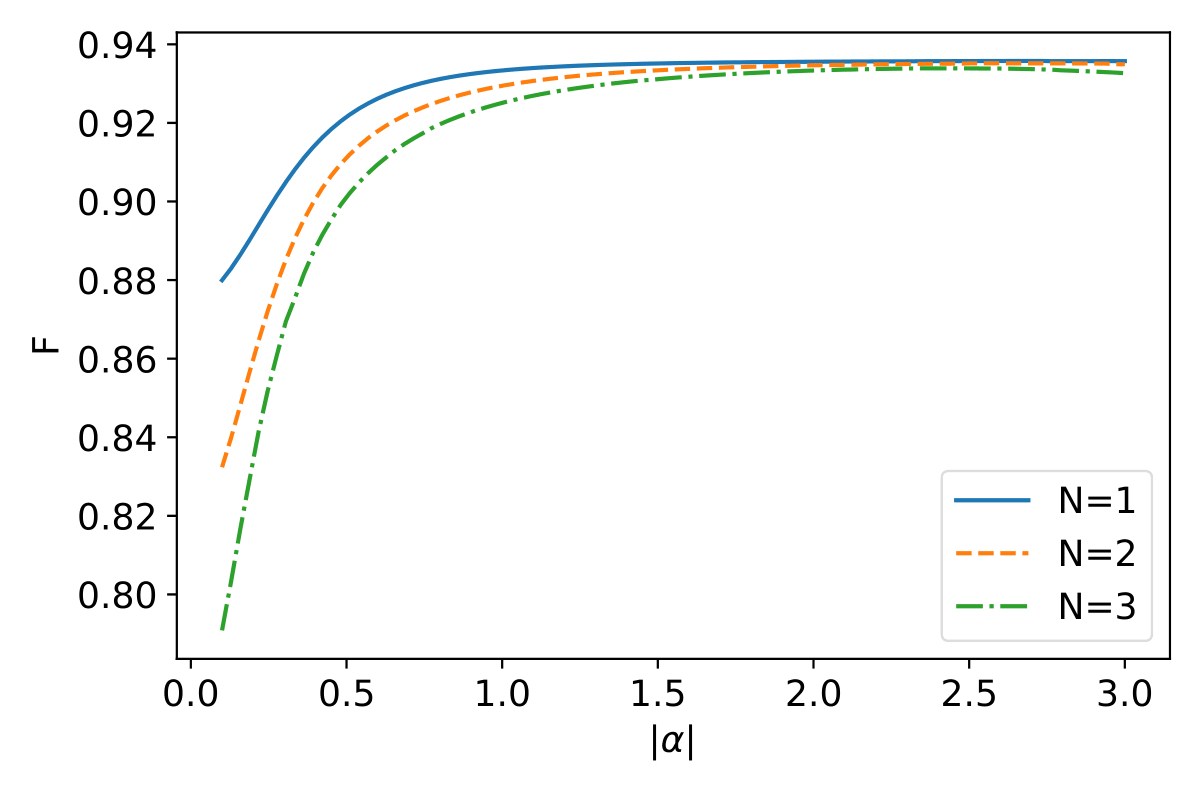}
	\caption{Fidelity as a function of $|\alpha|$ for $s = 0.5$,
		$\theta = \pi/4$, and $\phi - \beta = \pi/2$ rad. Here $\eta = 0.7$ and $\nu = 10^{-4}$.}
	\label{aaf}.
\end{figure}

\begin{figure}
	\centering
	\includegraphics[scale=0.7]{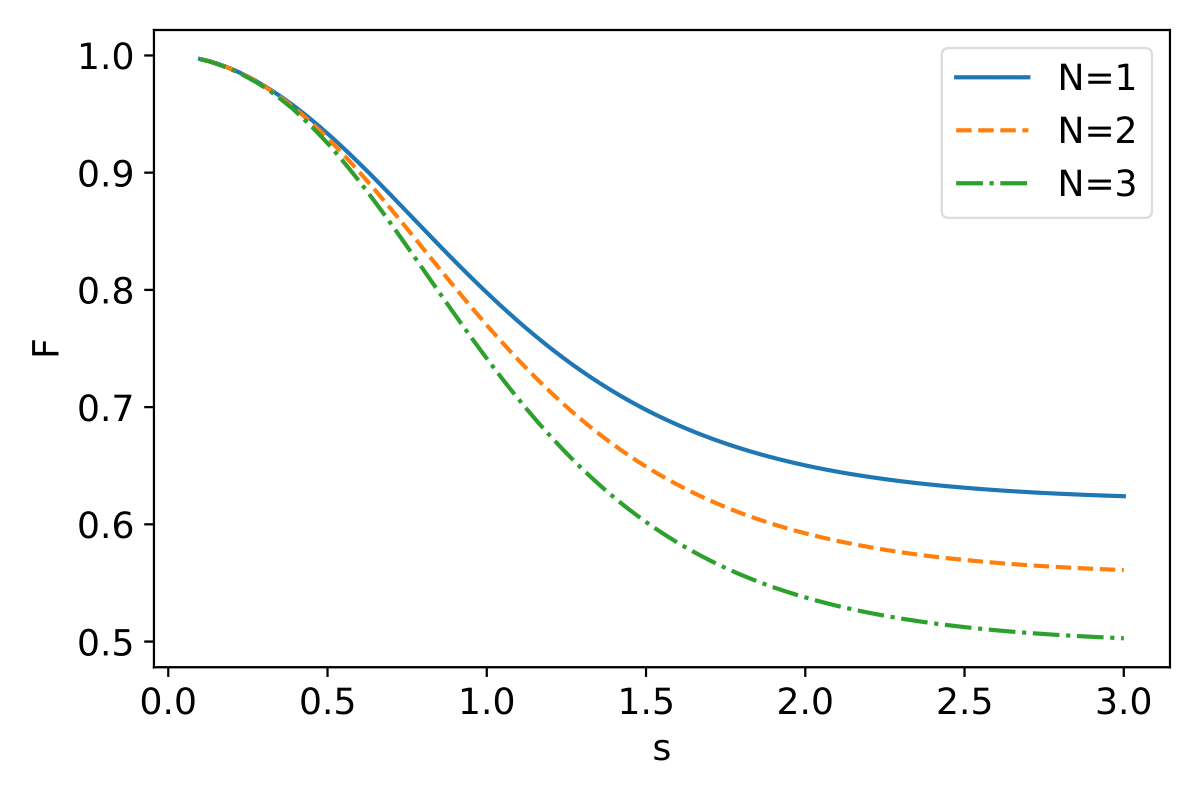}
	\caption{Fidelity as a function of $s$ for $|\alpha| = 1$, $\theta = \pi/4$ 
		and $\phi - \beta = \pi/2$ rad. Here $\eta = 0.7$ and $\nu = 10^{-4}$.}
	\label{asf}.
\end{figure}

The performance of the protocol may be assessed by calculating the fidelity $F$ of the output state in relation to state
$|\Phi^{(N,0)}_{a}\rangle$ (ideal output state), or
\begin{equation}
	F = \langle \Phi^{(N,0)}_{a}|\hat{\rho}^{(\eta,\nu)}_a|\Phi^{(N,0)}_{a}\rangle.
	\label{fidelity}
\end{equation}
We proceed by numerically computing the fidelity as a function of $|\alpha|$ and $s$, for different 
values of $N$, which is shown in Figs. \ref{aaf} and \ref{asf}, respectively. Firstly, we note that although 
the fidelity is clearly affected by the detection imperfections, it is possible to generate 
truncated states with  $F \gtrsim 0.9$. Thus, our modified scissors can, in principle, have a 
robustness against imperfections comparable to that of conventional scissors. Nevertheless, we observe 
that while in the conventional scissors (with coherent state input) the fidelity $F$ decreases with increasing 
$|\alpha|$ \cite{imoto01}, in our modified scissors $F$ increases with $|\alpha|$ instead, as seen in Fig. 
\ref{aaf}. On the other hand, the fidelity decreases with increasing pump strength $s$, as shown in Fig. 
\ref{asf}. This behavior can be understood if we take a closer look at the structure of the states generated 
by each type of scissors. Consider for simplicity the particular case of having $50:50$ beamsplitters and 
$N=1$. In a conventional scissors, the  
truncated state generated from a coherent state $|\alpha\rangle$ ($\alpha$ real) is of the 
form $|\varphi\rangle = {\cal N}(|0\rangle + \alpha |1\rangle)$, i.e., the coefficient of the one-photon state
is simply $\alpha$. In our modified scissors though, the resulting state is given by  
$|\varphi'\rangle = {\cal N}'(\alpha|0\rangle + \tanh s|1\rangle)$, and the coefficient of the one-photon 
state is $\tanh s$. Therefore, increasing the value of $\alpha$ $(s)$ in conventional (modified) scissors 
has the effect of decreasing the fidelity. Conversely, $\alpha$ is the coefficient of the vacuum 
state in the modified scissors output state and thus, increasing $\alpha$ should have the opposite 
effect in this case, that is, an increase of the fidelity. 

\section{Conclusion}\label{conclusion}

We proposed a scheme which allows quantum state truncation via the combined action 
of a nondegenerate optical parametric amplifier and a beamsplitter. This makes possible to 
perform the state truncation without the previous generation of Fock states.
In fact, there is no need of nonclassical input states whatsoever, and Gaussian states such as
vacuum states $+$ coherent states are sufficient resources to generate truncated output 
states which are nonclassical. This is clearly advantageous, given that the experimental
setup can be substantially simplified. We should point out that a single pumped nondegenerate 
parametric amplifier having vacuum states as inputs, generates a two-mode squeezed vacuum state 
as an output field. In our scheme, one of the modes of such an entangled state is 
mixed with an arbitrary field in a beamsplitter, and after the photodetections, the remaining
field mode is collapsed onto a truncated state. Accordingly, a nonclassical resource appropriate 
for state truncation is provided by the operation of the parametric amplifier itself, a device already 
integrating the proposed arrangement. 

Differently from the conventional scissors \cite{barnett98}, in our modified quantum scissors the 
nonclassical properties of the generated states can be selected by changing not only the transmittance of 
the beamsplitter, but also by adjusting the quantities associated to the classical pump in the amplifier, 
the strength $s$ and phase $\phi$. We should also point out that depending on the position of the
photodetectors, different classes of states can be produced. If the photodetectors are placed
in both output ports of the beamsplitter, a state having a maximum Fock number, say $N$, is
generated. However, if one photodetector is placed in one of the beamsplitter's port and the other in
the amplifier's output port, the generated state will have the Fock components $n < N$
removed. In other words, the scheme presented here allows the generation of states truncated in 
complementary sections of the Fock basis. Our results are expected to be relevant for exploring novel 
possibilities involving the combination of linear and nonlinear devices, aiming the manipulation of 
quantum states of light.

\section*{Acknowledgments}
 E.P.M. would like to acknowledge financial suport from CAPES (Coordenadoria de Aperfeiçoamento
 de Pessoal de Nível Superior), Brazil, grant N${\textsuperscript{\underline{o}}}$ 88887.514500/2020-00.
 This work was also supported by CNPq (Conselho Nacional para o
 Desenvolvimento Científico e Tecnológico, Brazil), via the INCT-IQ
 (National Institute for Science and Technology of Quantum Information),
 grant N${\textsuperscript{\underline{o}}}$ 465469/2014-0.


\section*{Appendix A: The derivation for Eq. (\ref{generalstate})}

\begin{figure}
	\centering
	\includegraphics[scale=1.25]{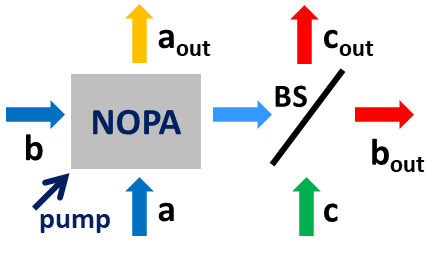}
	\caption{Schematic illustration of the proposed setup: a Nondegenerated Optical Parametric Amplifier (NOPA)
		with classical pump (strength $s$ and phase $\phi$), having $\hat{a}$ and $\hat{b}$ input modes.
		One of the output modes feeds a beamsplitter (BS), which
		has a second input mode, $\hat{c}$. Photodetectors may be placed in pairs in the output modes,
		either in $\hat{b}_{out}$ and $\hat{c}_{out}$ or $\hat{a}_{out}$ and $\hat{c}_{out}$.}
	\label{parampbs}
\end{figure}

We assume for the device in Fig. \ref{parampbs} a joint input state having vacuum states in the 
amplifier's input ports and an arbitrary pure state $|\psi\rangle = \sum \psi_i |i\rangle$ in the 
beamsplitter's input port. Thus, the combined action of the 
amplifier/beamsplitter $\hat{R}(\theta)\hat{S}(\xi)|\Psi_{in}\rangle$, will be
\begin{equation}
	|\Phi_{out}\rangle=\hat{R}(\theta)\hat{S}(\xi)|\Psi_{in}\rangle
	=\sum_{i=0}^\infty\frac{1}{\sqrt{i!}}\psi_i\hat{R}(\theta)\hat{S}(\xi)\hat{c}^{\dag i}
	|0, 0, 0\rangle.\label{initialstate}
\end{equation}
Using now the following relations,

\begin{eqnarray}
&&	\hat{S}\hat{a}^\dag\hat{S}^\dag=\hat{a}^\dag \cosh\,s+\hat{b}e^{-i\phi}\sinh\,s, \\ \nonumber
&&	\hat{S}\hat{b}^\dag\hat{S}^\dag=\hat{b}^\dag\cosh\,s+\hat{a}e^{-i\phi}\sinh\,s, \\ \nonumber
&&	\hat{S}\hat{c}^\dag\hat{S}^\dag=\hat{c}^\dag, \\ \nonumber
&&  \hat{S}|0, 0, 0\rangle=\sum_{n=0}^\infty \sech\,s(-e^{i\phi}\tanh\,s)^n |n, n, 0\rangle,
\end{eqnarray}

for the squeezing operator, and

\begin{eqnarray}
	&&	\hat{R}\hat{a}^\dag\hat{R}^\dag=\hat{a}^\dag, \\ \nonumber
	&&	\hat{R}\hat{b}^\dag\hat{R}^\dag=\mathtt{T}^*\hat{b}^\dag+\mathtt{R}\hat{c}^\dag, \\ \nonumber
	&&	\hat{R}\hat{c}^\dag\hat{R}^\dag=-\mathtt{R}^*\hat{b}^\dag+\mathtt{T}\hat{c}^\dag, \\ \nonumber
	&&	\hat{R}|0, 0, 0\rangle=|0, 0, 0\rangle
\end{eqnarray}

for the beamsplitter operator.

Applying the relations above to Eq. (\ref{initialstate}), we obtain

\begin{eqnarray}
	|\Phi_{out}\rangle&=&\hat{R}(\theta)\hat{S}(\xi)|\Psi_{in}\rangle=
	\sum_{i=0}^\infty\sum_{n=0}^\infty\frac{1}{\sqrt{i!}}\psi_iA_n(\xi)\hat{R}\hat{c}^{\dag i}|n, n, 0\rangle \nonumber \\
	&=&\sum_{i=0}^\infty\sum_{n=0}^\infty\frac{1}{\sqrt{i!}}\frac{1}{\sqrt{n!}}\psi_iA_n(\xi)\hat{R}\hat{b}^{\dag n}\hat{c}^{\dag i}|n, 0, 0\rangle \\ \nonumber
	&=&\sum_{i=0}^\infty\sum_{n=0}^\infty\frac{1}{\sqrt{i!}}\frac{1}{\sqrt{n!}}\psi_i\, 
	A_n(s,\phi)(\mathtt{T}^*\hat{b}^\dag+\mathtt{R}\hat{c}^\dag)^n(-\mathtt{R}^*\hat{b}^\dag+
	\mathtt{T}\hat{c}^\dag)^i|n, 0, 0\rangle,
\end{eqnarray}
which results in
\begin{equation}
	|\Phi_{out}\rangle=\sum_{i=0}^\infty\sum_{n=0}^\infty\sum_{j=0}^i\sum_{m=0}^n\frac{1}{\sqrt{i!}}\frac{1}{\sqrt{n!}}
	\frac{i!}{j!(i-j)!}\frac{n!}{m!(n-m)!}\psi_i A_n(s,\phi)  
	(\mathtt{T}^*\hat{b}^\dag)^{n-m}(\mathtt{R}\hat{c}^\dag)^m
	(-\mathtt{R}^*\hat{b}^\dag)^{i-j}(\mathtt{T}\hat{c}^\dag)^j|n, 0, 0\rangle.
\end{equation}
Finally,
\begin{eqnarray}
	|\Phi_{out}\rangle&=&\sum_{i=0}^\infty\sum_{n=0}^\infty\sum_{j=0}^i\sum_{m=0}^n\frac{\sqrt{i!}}{j!(i-j)!}
	\frac{\sqrt{n!}}{m!(n-m)!} 
	\sqrt{(i-j+n-m)!} \sqrt{(j+m)!}\,\psi_i\, A_n(s,\phi) \nonumber \\
	&\times& \mathtt{T}^j \mathtt{T}^{*n-m} 
	\mathtt{R}^m(-\mathtt{R}^*)^{i-j}|n, i-j+n-m, j+m\rangle. 
\end{eqnarray}

\newpage

\bibliographystyle{unsrt}
\bibliography{refsscissors}
\newpage

\end{document}